\documentclass[prb,reprint,superscriptaddress,floatfix,amsmath,amssymb]{revtex4-2}
\usepackage{color}
\usepackage{graphicx}
\usepackage{float}
\usepackage{hyperref}
\usepackage[maxfloats=256]{morefloats}
\usepackage{comment}
\usepackage[small, bf]{caption}
\usepackage[subrefformat=parens]{subcaption}
\usepackage[dvipsnames]{xcolor}
\usepackage{xspace} 
\usepackage{ulem} 

\usepackage[version=3]{mhchem}
\usepackage{mathrsfs}
\usepackage{bm}
\usepackage{physics}
\hypersetup{colorlinks=true,urlcolor=blue,citecolor=blue,linkcolor=blue,breaklinks=true}

\newcommand\ee{\mathrm{e}}
\newcommand\ii{\mathrm{i}}
\newcommand\Sigmapred{\Sigma_\mathrm{pred}}
\newcommand\Sigmatest{\Sigma_\mathrm{test}}
\newcommand\w{\omega}
\newcommand\iw{{\mathrm{i}\omega}}
\newcommand\ibarw{\mathrm{i}\bar{\omega}}
\newcommand\itildew{\mathrm{i}\tilde{\omega}}
\newcommand\barw{\bar{\omega}}
\newcommand\wmax{\ensuremath{{\omega_\mathrm{max}}}}
\newcommand\bk{\ensuremath{{\boldsymbol{k}}}}

\begin{document}

\allowdisplaybreaks
\title{Physics-informed neural network model for quantum impurity problems based on Lehmann representation}

\author{Fumiya Kakizawa}
\affiliation{Department of Physics, Saitama University, Saitama 338-8570, Japan}

\author{Satoshi Terasaki}
\affiliation{AtelierArith}

\author{Hiroshi Shinaoka}
\affiliation{Department of Physics, Saitama University, Saitama 338-8570, Japan}

\begin{abstract}
We propose a physics-informed neural network (PINN) model to efficiently predict the self-energy of Anderson impurity models (AIMs) based on the Lehmann representation.
As an example, we apply the PINN model to a single-orbital AIM (SAIM) for a noninteracting electron bath with a semicircular density of states.
Trained across a wide range of onsite Coulomb interactions $U$ and hybridization strengths $V$, the PINN model demonstrates high accuracy in both $U$-$V$ and Matsubara-frequency spaces. 
Additionally, we investigate the effectiveness of physical constraints implemented in the PINN model.
For example, We show that the Lehmann representation allows the PINN model to reduce the maximum test error in an electron filling by a factor of approximately 7.8.
\end{abstract}

\maketitle

\section{Introduction}\label{sec:introduction}
Dynamical mean-field theory (DMFT)~\cite{RevModPhys.68.13} is a powerful tool for studying strongly correlated electron systems.
DMFT is based on mapping an original lattice model onto a quantum Anderson impurity model (AIM)~\cite{PhysRev.124.41}, involving a limited number of local quantum degrees of freedom embedded in a bath determined by a self-consistent condition.
Several accurate numerical methods (\textit{impurity solvers}) have been proposed. 
These include exact diagonalization (ED) methods~\cite{PhysRevLett.72.1545},
continuous-time quantum Monte Carlo methods~\cite{PhysRevB.72.035122, PhysRevLett.97.076405},
numerical renormalization group method~\cite{SAKAI1994307},
tensor network methods~\cite{PhysRevLett.93.246403, PhysRevX.7.031013, PhysRevB.104.115119}
and tensor cross interpolation method~\cite{PhysRevX.12.041018} among others.
However, these impurity solvers are computationally demanding and must be used iteratively within the DMFT calculations, making the development of more efficient impurity solvers an important research subject.

This high computational cost gave rise to a pressing need to develop cost-effective predictors that can accurately predict the calculation results of the more accurate existing impurity solvers.
Machine learning (ML) models are promising candidates for such predictors. 
Numerous ML models have been developed to predict the Green's function~\cite{PhysRevB.90.155136, PhysRevB.104.205120, zhao2023languageinspiredmachinelearningapproach, agapov2024predictinginteractinggreensfunctions} or its spectral function~\cite{PhysRevB.103.245118, Ren_2021} specifically for the single-orbital AIM (SAIM).
However, for DMFT calculations, the primary requirement is the self-energy rather than the Green's function. 
Since deriving the self-energy from Green's function via the Dyson equation can be numerically unstable, it is essential to develop ML models that can directly predict the self-energy.

In general, it is challenging for a simple ML model to estimate self-energy accurately, even for SAIM.
This difficulty arises from the crossover behavior of the SAIM between \textit{itinerant} and \textit{localized} regimes, leading to significant quantitative differences in the self-energy between them.
Notably, the absolute values of the self-energy differ by orders of magnitude between them. 
Furthermore, at low Matsubara frequencies, the qualitative behavior of the self-energy also  varies dramatically: 
its imaginary part vanishes linearly in the itinerant regime and grows in the localized regime.

To overcome this difficulty, we propose a physics-informed neural network (PINN)~\cite{raissi2017physicsinformeddeeplearning1, raissi2017physicsinformeddeeplearning2} model, which combines deep learning techniques with Lehmann representation~\cite{PhysRevB.109.075112} and sparse modeling techniques~\cite{PhysRevB.96.035147, PhysRevB.101.035144}, specifically designed to predict the self-energy of the SAIM.
We demonstrate that this PINN model effectively estimates the self-energy in both itinerant and localized regimes and provides a significant advantage over non-physics-informed neural network models.

This paper is organized as follows: 
Section~\ref{sec:model} introduces the SAIM. 
In Sec.~\ref{sec:nn}, we introduce the sparse grids in Matsubara-frequency and real-frequency spaces and the PINN model.
In Sec.~\ref{sec:training}, we describe the overall procedure for calculating the self-energy using ED and detail the training conditions.
In Sec.~\ref{sec:results}, we present the numerical results of the predicted self-energy and other physical quantities derived from it and highlight the advantages over non-physics-informed neural network models.

\section{Single-orbital Anderson impurity model}\label{sec:model}
\begin{figure}
    \centering
    \includegraphics[width=0.9\linewidth]{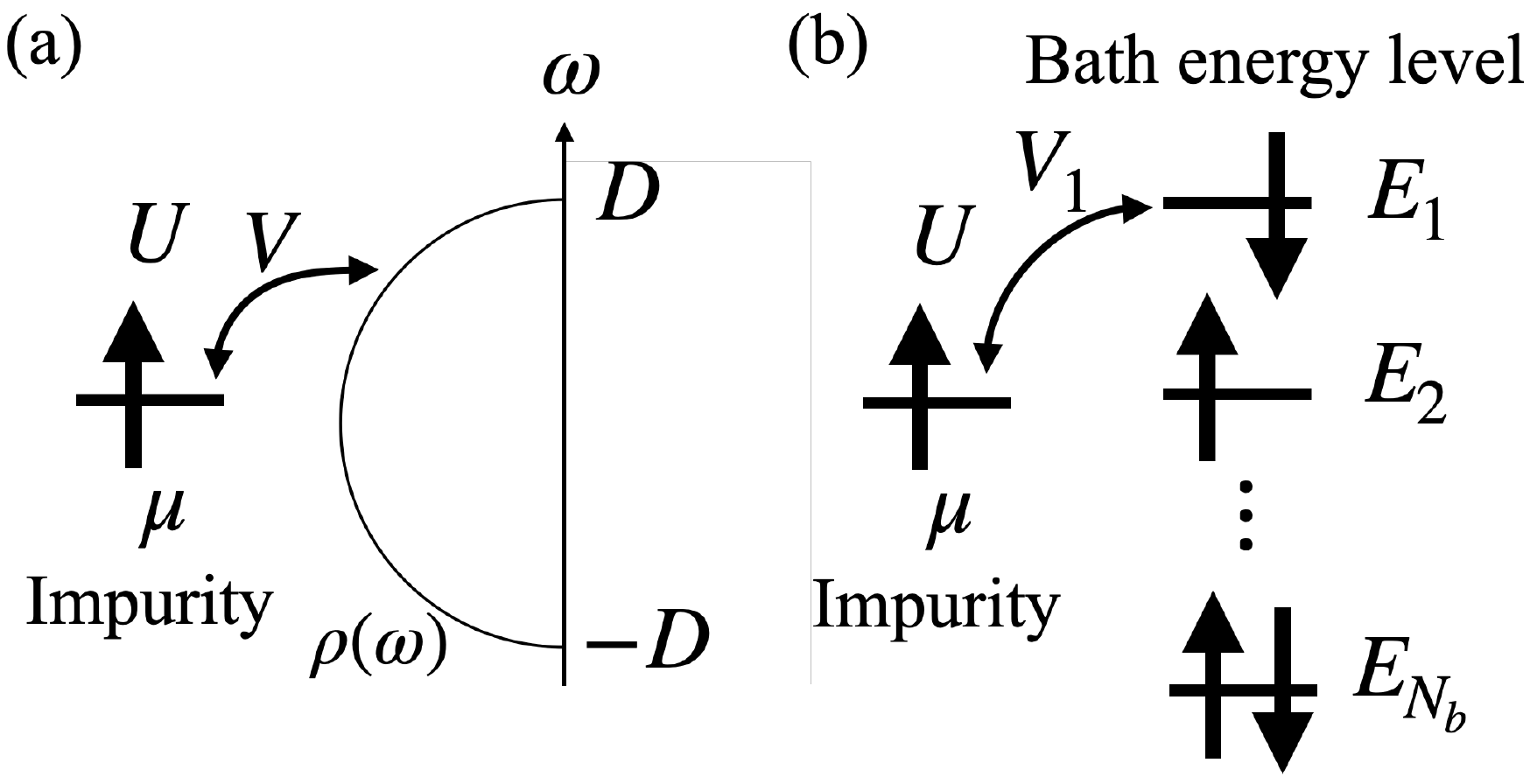}
    \caption{SAIMs.
            (a) An impurity hybridizes with a bath with semicircular density-of-states.
            (b) An impurity hybridizes with discretized bath sites arranged in a star-like geometry.
    }
    \label{fig:AIM}
\end{figure}

We consider the SAIM depicted in Fig.~\ref{fig:AIM}(a). 
This model consists of a single localized impurity orbital (with two-fold spin degeneracy) hybridizing with a noninteracting bath.
Its Hamiltonian is given by
\begin{align}
   H_{\mathrm{SAIM}} &= \sum_{\bk,\sigma} \epsilon_\bk c^\dagger_{\bk\sigma} c_{\bk\sigma}  + \mu \sum_\sigma n_\sigma + U n_\uparrow n_\downarrow \nonumber\\
  & + V \sum_{\bk,\sigma} (c^\dagger_{\bk,\sigma} d_\sigma + \mathrm{H.c.}),
  \label{eq:SAIM}
\end{align}
where $\epsilon_\bk$ is the bath dispersion with bath level $\bk$, $c_{\bk\sigma}$ and $c^\dagger_{\bk\sigma}$ are the annihilation and creation operators for electrons for $\bk$ and local spin $\sigma$. 
Furthermore, the operators $d_\sigma, d^\dagger_\sigma$, and $n_\sigma = d^\dagger_\sigma d_\sigma$ are the annihilation, creation, and number operators for electrons with $\sigma$ on the impurity, respectively.
Additionally, $U$ is the Coulomb repulsion between electrons on the impurity, and $\mu$ is the chemical potential that regulates the electron occupancy on the impurity.
The second line of Eq.~\eqref{eq:SAIM} represents the hybridization between the impurity and the bath with coupling strength $V$.
We further consider a noninteracting bath with semicircular density-of-states (DOS) $\rho(\w) = \frac{2}{\pi D}\sqrt{1 - (\w/D)^2}$ with a half-bandwidth $D=2$.

The impurity Green's function is defined as follows: 
\begin{align}
   G_{\sigma\sigma'}(\tau) = - \expval{\mathcal{T}_\tau d_{\sigma}(\tau)d^\dagger_{\sigma'}},
   \label{eq:green_func}
\end{align}
where $\mathcal{T_\tau}$ is the imaginary time-ordering operator and $\expval{A}$ represents the expectation value of an operator $A$, defined as $\expval{A}\equiv \frac{\Tr A\ee^{-\beta H_{\mathrm{SAIM}}}}{\Tr\ee^{-\beta H_{\mathrm{SAIM}}}}$ with $\beta$ being the inverse temperature.
We set $\hbar = 1$ and $k_{\mathrm{B}}=1$.
Here, $A(\tau)$ is a Heisenberg representation of an operator $A$, defined as $A(\tau) = \ee^{\tau H_{\mathrm{SAIM}}}A\ee^{-\tau H_{\mathrm{SAIM}}}$.
In this study, we ignore the spin dependence and focus on paramagnetic cases.
The Fourier transformation of $G(\tau)$ is given by
\begin{align}
   G(\iw_n) \equiv \int_{0}^{\beta}\dd\tau G(\tau)\ee^{\iw_n\tau},
\end{align}
where $\iw_n = \ii(2n+1)\pi/\beta$ is a fermionic Matsubara frequency ($n$ is an integer).

The self-energy $\Sigma(\iw)$ is defined by the local Dyson equation 
\begin{align}
   \Sigma(\iw) = \iw - \mu - \Delta(\iw)- G^{-1}(\iw), 
   \label{eq:dyson_eq}
\end{align}
where $\Delta(\iw)$ is the hybridization function parameterizing the bath given by
\begin{align}
    \Delta(\iw) &= \sum_{\bk}\frac{V^2}{\iw - \epsilon_\bk}.
    \label{eq:hybfunc_cont}
\end{align}
Different sets of $V$ and $\epsilon_\bk$ can yield the same $\Delta(\iw)$ numerically. 
This suggests that when constructing an ML model, it is more effective to parameterize the bath using $\Delta(\iw)$ rather than $V$ and $\epsilon_\bk$. 

Here, we briefly describe the properties of the SAIM in the localized and itinerant regimes.
In the localized regime, where $U$ is large and $V$ is small, the SAIM exhibits a local moment on the impurity, with the absolute value of the imaginary part of the self-energy growing as the Matsubara frequency approaches zero.
As $U$ and $V$ are adjusted, the SAIM undergoes a crossover from the localized regime to the itinerant regime~\cite{PhysRevLett.115.236601}, where the bath screens the local moment of the impurity.
In the itinerant regime, the imaginary part of the self-energy linearly vanishes at low Matsubara frequencies.

\section{Physics-informed neural network model}\label{sec:nn}
In this section, we introduce our PINN model.
In Sec.~\ref{sec:sparse_grids}, we introduce the sparse grids in Matsubara- and real-frequency spaces based on the sparse-modeling techniques.
Using these grids, we define the PINN model in Sec.~\ref{subsec:nn}.

\subsection{Sparse grids in Matsubara- and real-frequency spaces based on the Lehmann representation}\label{sec:sparse_grids}
    
In this subsection, we introduce the sparse Matsubara-frequency grid~\cite{PhysRevB.101.035144} and real-frequency grid~\cite{10.21468/SciPostPhysLectNotes.63} generated by the so-called intermediate representation (IR)~\cite{PhysRevB.96.035147}, which we review in the following.
The Lehmann representation of the imaginary-time Green's function is given by
\begin{align}
G(\tau) &= - \int_{-\wmax}^{\wmax} \dd{\w} K(\tau, \w) \rho(\w),\label{eq:lehmann}
\end{align}
where $\rho(\w)$ is the spectral function associated with $G(\tau)$ and we set $0 < \tau < \beta$.
For fermionic Green's functions, the kernel is given by
\begin{align}
    K(\tau, \w) = \frac{\ee^{-\tau \w}}{1 + \ee^{-\beta\w}}.
\end{align}
In the Matsubara-frequency space, the Lehmann representation reads
\begin{align}
G(\iw_n) &= \int_{-\wmax}^{\wmax} \dd{\w} K(\iw_n, \w) \rho(\w),\label{eq:lehmann-matsubara}
\end{align}
where
\begin{align}
K(\iw_n, \w) &= \frac{1}{\iw_n - \w}.
\end{align}

The self-energy and hybridization functions have the same mathematical structure as Green's functions.
For instance, the Lehmann representation of the self-energy is given by
\begin{align}
\Sigma(\iw_n) = \int_{-\wmax}^{\wmax}\dd{\w}\frac{\rho_{\mathrm{aux}}(\w)}{\iw_n - \w},
\end{align}
where $\rho_{\mathrm{aux}}(\w)$ is an auxiliary spectral function associated with the self-energy.

Because the kernel is ill-conditioned, $G(\tau)$ and $G(\iw)$ have less information than $\rho(\w)$~\cite{PhysRevB.96.035147}.
This allows us to construct an optimal compact orthogonal basis, the IR basis, for imaginary-time objects through the singular value expansion:
\begin{align}
    K(\tau, \w) = \sum_{l=0}^{\infty}U_l(\tau)S_lV_l(\w),
\end{align}
where $S_l$ is the exponentially decaying singular value, and $U_l(\tau)$ and $V_l(\w)$ are the singular functions parameterized by the dimensionless parameter $\Lambda \equiv \beta \wmax$.
The sets of the singular functions $\{U_l(\tau)\}$ and $\{V_l(\w)\}$ serve as the IR bases in imaginary-time and real-frequency spaces, respectively.
The number of nonzero singular values $N_\mathrm{IR}$ increases only logarithmically with $\Lambda$~\cite{Chikano18}.

One can expand any Green's function in the IR basis as
\begin{align}
G(\tau) &\simeq \sum_{l=0}^{N_\mathrm{IR}-1} G_l U_l(\tau),
\end{align}
where the expansion coefficients $G_l~(\equiv S_l \int_{-\wmax}^{\wmax} \dd{\w} \rho(\w) V_l(\w))$ are guaranteed to decay as fast as $S_l$.
The IR basis functions in the Matsubara-frequency space $\{U_l(\iw)\}$ are defined by Fourier transforming $\{U_l(\tau)\}$.

\begin{figure}
    \centering
    \includegraphics[width=0.8\linewidth]{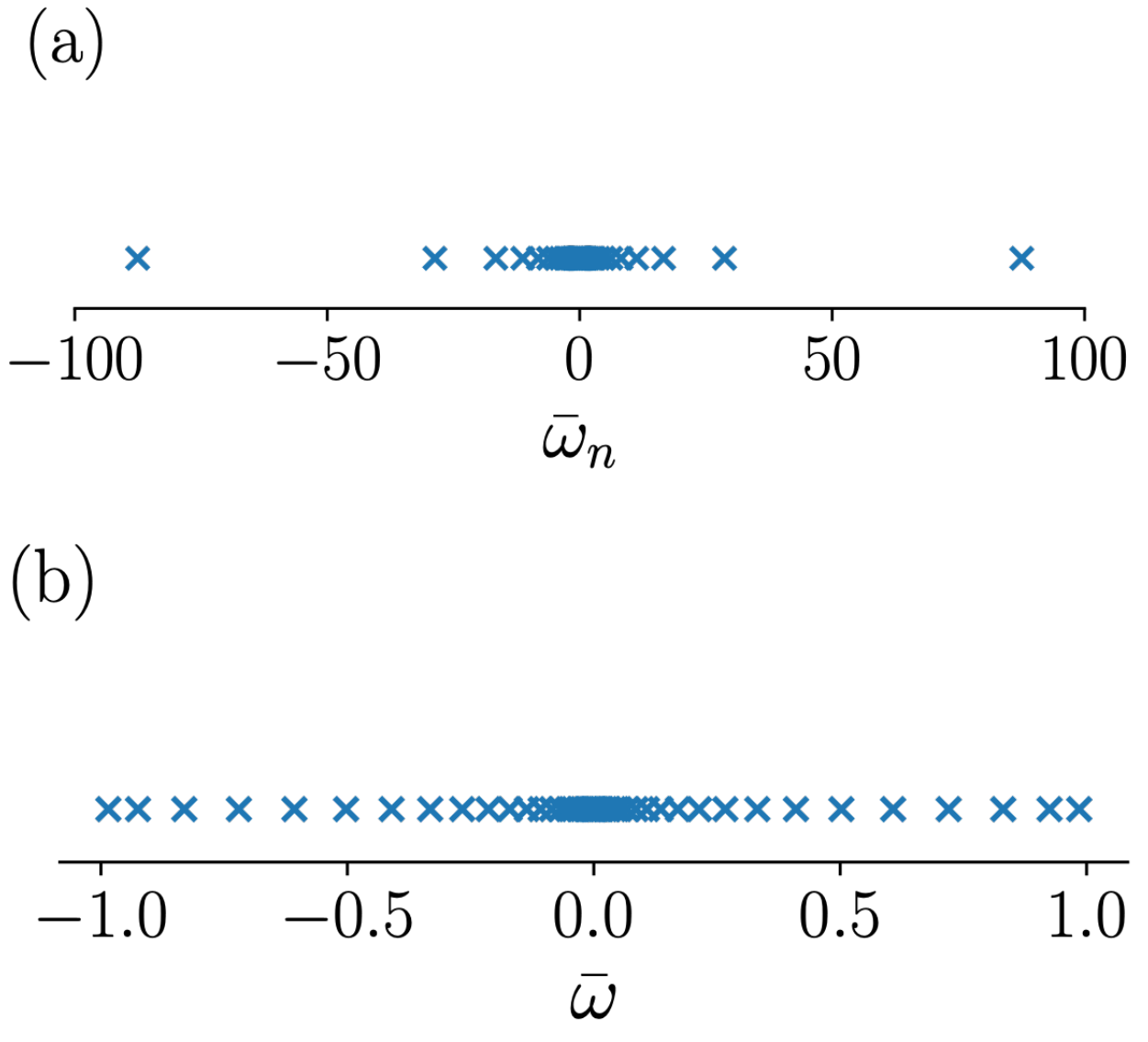}
    \caption{Sparse grid in (a) Matsubara-frequency and (b) real-frequency spaces for $\Lambda=1000$.
    }
    \label{fig:sparse_grid}
\end{figure}

One can also construct a sparse grid in Matsubara-frequency space, denoted as $\{\ibarw_i\}$~\cite{PhysRevB.101.035144}, and a sparse grid in real-frequency space, denoted as $\{\bar{\w}_i\}$~\cite{10.21468/SciPostPhysLectNotes.63}, both associated with the IR basis.
Figures~\ref{fig:sparse_grid}(a) and~\ref{fig:sparse_grid}(b) show the sparse Matsubara-frequency and real-frequency grids for $\Lambda=1000$.

These grids have the following favorable properties.
First, the grid sizes ($\simeq N_\mathrm{IR}$) grow logarithmically with $\Lambda$ as the IR basis.
Second, the IR expansion coefficients (e.g., $G_l$) of a Green's-function-like object can be reconstructed in a numerically stable way from its numerical values known on the Matsubara-frequency sparse grid $\{\ibarw_i\}$.
In other words, the Matsubara-frequency sparse grid can capture the full information on the Matsubara-frequency dependence.
Third, one can parameterize the Matsubara-frequency dependence of any Green's-function-like object via the Discrete Lehmann Representation (DLR)~\cite{PhysRevB.105.235115} as
\begin{align}
G(\iw_n) &\simeq \sum_i \frac{c_i}{\iw_n - \bar{\w}_i},
\end{align}
using the real-frequency grid $\{\bar{\w}_i\}$.
We refer the reader for more details on the sparse representations to Ref.~[\onlinecite{PhysRevB.96.035147}].

\subsection{Network structure}\label{subsec:nn}
\begin{figure*}
    \centering
    \includegraphics[width=1\linewidth]{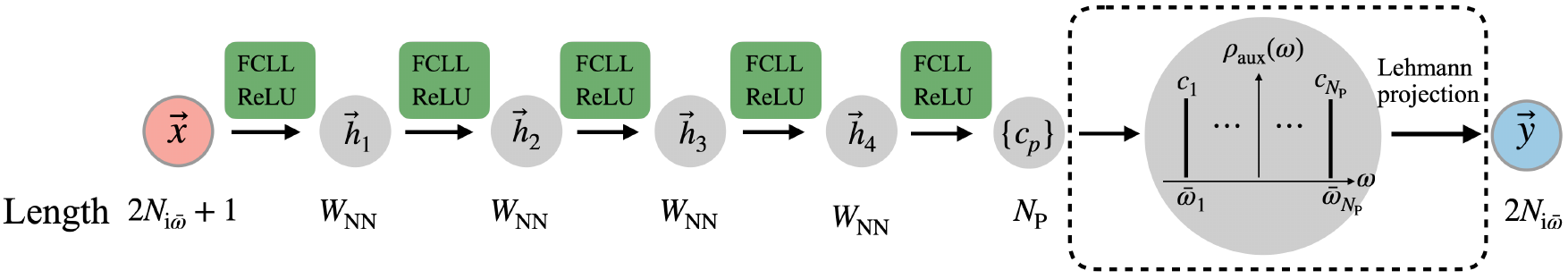}
    \caption{Schematic of the PINN model:
     The input layer, indicated by a red circle, processes the input vector $\vec{x}$, which consists of $U$ and the sequence of real and imaginary parts of $\Delta(\ibarw_n)$.
     The sequence of green squares represents hidden layers, including the FCLL and the ReLU activation function, where $\vec{h}_i$ denotes the output vector from the $i$th hidden layer. 
     In the Lehmann layer, depicted as a square with black dashed lines, the Lehmann projection $\hat{L}$ defined in Eq.~\eqref{eq:lehmann_proj} is applied to $\{c_p\}$.
     The output layer, shown as a blue circle, yields the output vector $\vec{y}$, which consists of the sequence of real and imaginary parts of $\Sigma(\ibarw_n)$.
     The architecture parameters include the number of sparse Matsubara-frequency grid points $N_{\ibarw}$, the number of poles $N_{\mathrm{P}}$, the width of the first to fourth hidden layers $W_{\mathrm{NN}}$. 
    }         
    \label{fig:NNmodel}
\end{figure*}

This subsection briefly reviews the structure of the PINN model. 
Figure~\ref{fig:NNmodel} illustrates the PINN model for predicting the self-energy $\Sigma(\iw)$, taking inputs $U$ and $\Delta(\iw)$.
All data in the Matsubara-frequency space and real-frequency spaces are represented on the sparse grids.
The model consists of the input layer (a red circle), five hidden layers comprised only of the standard fully connected linear layer (FCLL) and the rectified linear unit (ReLU) activation function ($\mathrm{ReLU}(x) = \max(0, x)$) (green squares), the Lehmann layer applying a Lehmann projection (a square with black dashed lines), and the output layer (a blue circle). 

The input vector $\vec{x}$ is given by
\begin{align}
    \vec{x}=(U, &\Re \Delta(\ibarw_1), \Re \Delta(\ibarw_2), \cdots, \Re \Delta(\ibarw_{N_{\ibarw}}), \\ \nonumber
                &\Im \Delta(\ibarw_1), \Im \Delta(\ibarw_2), \cdots, \Im \Delta(\ibarw_{N_{\ibarw}})),
\end{align}
where $N_{\ibarw}$ indicates the total number of the sparse Matsubara-frequency grid points.
Splitting the real and imaginary parts of $\Delta(\ibarw)$ ensures that the NN model parameters are real numbers, allowing the ReLU activation function to operate.

In the hidden layers, the input vector is transformed linearly using the weight matrix and the bias vector, followed by the application of the ReLU activation function. 
The width of the first to fourth hidden layers, denoted as $W_{\mathrm{NN}}$, is set to $1696$.
As shown in Appendix~\ref{sec:width_dep}, this suffices to obtain converged results.
In the fifth hidden layer, the width is set to the number of sampling points (\textit{poles}) in the real-frequency space, totaling 424.
These sampling points are generated by dividing the space between the sparse real-frequency grid points into equal intervals by a factor of 10~\footnote{
For instance, the DLR grid is designed to describe the imaginary propagator without the non-negative constraint accurately.
For instance, the contribution of a pole with a positive coefficient between two adjacent poles requires positive and negative weights on neighboring DLR grid points.
The oversampling alleviates this problem.
}.
We confirmed that the results shown below are converged with respect to the number of sampling points in the real-frequency space.

In the Lehmann layer, the self-energy $\Sigma(\ibarw)$ is parameterized by using a pole representation of $\rho_{\mathrm{aux}}(\w)$ as
\begin{align}
    \rho_\mathrm{aux}(\w) &= \sum_{p=1}^{N_{\mathrm{P}}} c_p \delta(\w - \barw_p),\\
    \Sigma(\ibarw_n) &= \int_{-\wmax}^{\wmax} \dd{\w}\frac{\rho_\mathrm{aux}(\w)}{\ibarw_n - \w} = \sum_{p=1}^{N_{\mathrm{P}}} \frac{c_p}{\ibarw_n - \barw_p}, 
\end{align}
where $c_p \ge 0$ and $N_{\mathrm{P}}$ denotes the number of poles.
In our PINN model, the application of the ReLU activation function in the fifth hidden layer, which we refer to as the non-negative projection, ensures $c_p \ge 0$ for all $p$.
This parameterization with the nonnegative $\rho_\mathrm{aux}(\w)$ ensures that the output satisfies various important properties of exact $\Sigma(\iw)$, e.g., $\frac{\partial^2 \Sigma(\tau)}{\partial \tau^2} \le 0$, where $\Sigma(\tau)$ is the Fourier transformation of $\Sigma(\iw)$ defined by $\Sigma(\tau)=\frac{1}{\beta}\sum_{n=-\infty}^{\infty}\ee^{-\iw_n\tau}\Sigma(\iw_n)$.
The Lehmann projection is defined as follows:
\begin{align}
    \hat{L} = 
   \left(
   \begin{matrix}
   \mathrm{Re}\frac{1}{\ibarw_1-\barw_1}  & \cdots  & \mathrm{Re}\frac{1}{\ibarw_1-\barw_{N_{\mathrm{P}}}} \\
   \vdots & \ddots & \vdots \\ 
   \mathrm{Re}\frac{1}{\ibarw_{N_{\ibarw}}-\barw_1} & \cdots & \mathrm{Re}\frac{1}{\ibarw_{N_{\ibarw}}-\barw_{N_{\mathrm{P}}}}\\
   \\ 
   \mathrm{Im}\frac{1}{\ibarw_1-\barw_1}  & \cdots  & \mathrm{Im}\frac{1}{\ibarw_1-\barw_{N_{\mathrm{P}}}} \\
   \vdots & \ddots & \vdots \\ 
   \mathrm{Im}\frac{1}{\ibarw_{N_{\ibarw}}-\barw_1} & \cdots & \mathrm{Im}\frac{1}{\ibarw_{N_{\ibarw}}-\barw_{N_{\mathrm{P}}}}\\
   \end{matrix}
   \right).
   \label{eq:lehmann_proj}
\end{align}
We note that the Lehmann projection is not learned through backpropagation; rather, it is established beforehand during the construction of the Matsubara-frequency and real-frequency sparse grids.

The output vector $\vec{y}$ is given by
\begin{align}
    \vec{y}=(&\Re \Sigma(\ibarw_1), \Re \Sigma(\ibarw_2), \cdots, \Re \Sigma(\ibarw_{N_{\ibarw}}), \\ \nonumber
             &\Im \Sigma(\ibarw_1), \Im \Sigma(\ibarw_2). \cdots, \Im \Sigma(\ibarw_{N_{\ibarw}})).
\end{align}

A notable point is that the PINN model does not directly read the bare bath parameters and depends on these parameters only through $\Delta(\itildew)$, which may increase model versatility.

\section{Training method}\label{sec:training}
In this section, we explain the training method. 
In Sec.~\ref{sec:data_generation}, we outline the procedure for calculating the training data for the self-energy using the ED method.
In Sec.~\ref{sec:training_condtions}, we describe the other training conditions. 

\subsection{Calculation of the self energy using the exact diagonalization method}\label{sec:data_generation}
In this subsection, we describe the calculation of the self-energy $\Sigma(\iw)$ using the ED method.
We first discretize the model (\ref{eq:SAIM}) with a finite number of auxiliary bath sites as 
\begin{align}
   H_{\mathrm{SAIM}} &= \sum_{b,\sigma}E_b c^\dagger_{b\sigma} c_{b\sigma} + \mu \sum_\sigma n_\sigma + U n_\uparrow n_\downarrow\nonumber\\
  & + \sum_{b,\sigma}^{N_b} (V_b c^\dagger_{b,\sigma} d_\sigma + \mathrm{H.c.}),
\end{align}
such that the hybridization function $\tilde \Delta(\iw) = \sum_{b=1}^{N_\mathrm{b}} \frac{V_b^2}{\iw - E_b}$ approximates the original $\Delta(\iw)$ on the Matsubara-frequency axis accurately. 
In practice, we use non-linear fitting to determine bath parameters $\{V_b\}$ and $\{E_b\}$.
Here, $N_b$ indicates the number of bath sites, $\epsilon_k$ is replaced with $E_b$, and $V_b$ denotes the coupling strength between the impurity and the $b$th bath site.
Figure~\ref{fig:AIM}(b) depicts the discretized SAIM with a star-like bath geometry where the bath sites are exclusively connected through the impurity.
Although the continuous limit can be exactly recovered only when $N_b$ is infinite, we confirmed that $N_b=6$ is sufficient to provide a good approximation.

We then diagonalize this finite-size Hamiltonian, which includes $N_b+1$ sites, using either the full diagonalization method or the Lanczos method.
Subsequently, we calculate $G(\tau)$ defined in Eq.~\eqref{eq:green_func}, followed by Fourier transforming $G(\tau)$ to $G(\iw)$.
Finally, $\Sigma(\iw)$ is obtained by solving the Dyson equation (\ref{eq:dyson_eq}).
These calculations are performed on the imaginary-time and Matsubara-frequency sparse grids~\cite{PhysRevB.101.035144}.

\begin{figure}
    \centering
    \includegraphics[width=1\linewidth]{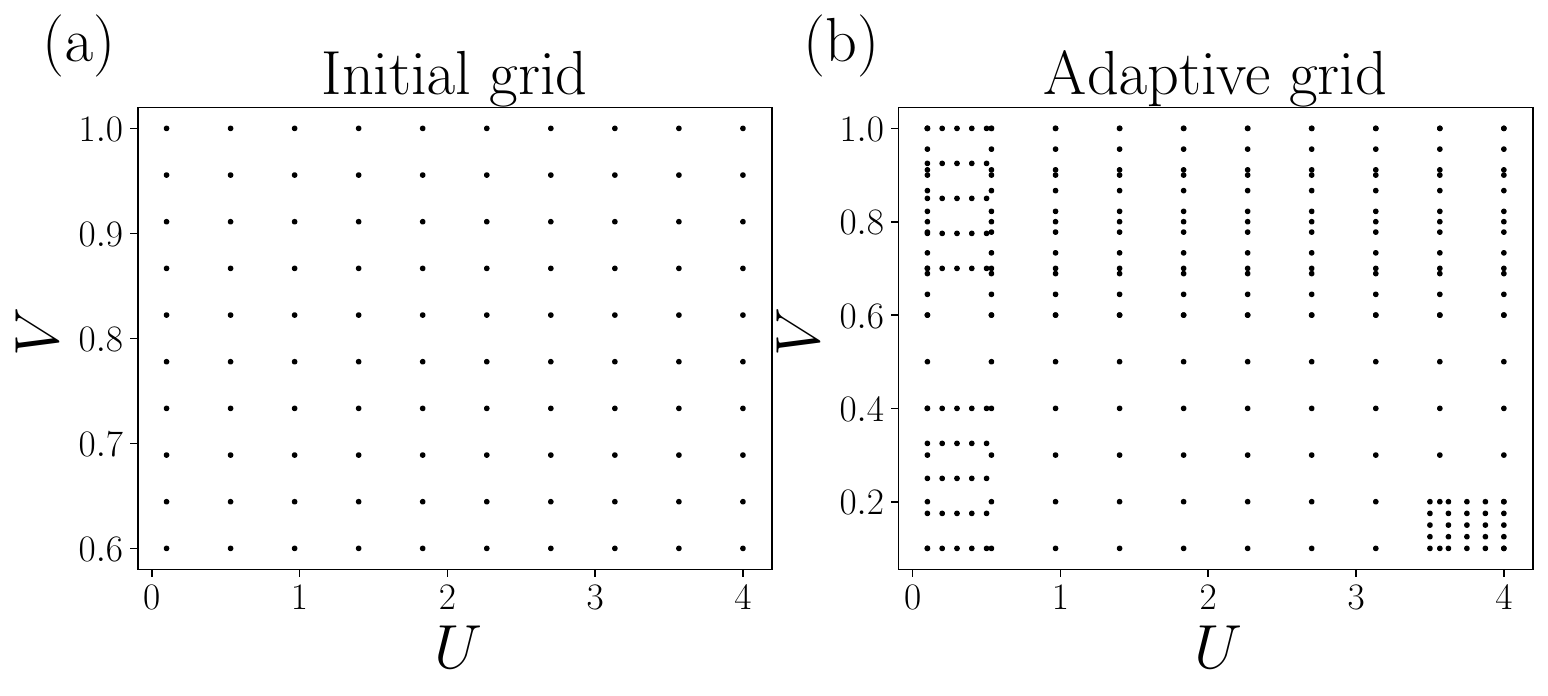}
    \caption{Training data distributions in $U$-$V$ space: 
             (a) the initial equidistant grid totaling 100 samples, 
             (b) the adaptive grid totals 275 samples.
    }
    \label{fig:training_grid}
\end{figure}

In this study, we have generated the training data for various values of $U$ and $V$.
The ranges of $U$ and $V$ are 0.1--4.0 and 0.1--1.0, respectively.
Additionally, we set $\mu = - U/2$ to ensure particle-hole symmetry.
We first conducted training on an equidistant grid shown in Fig.~\ref{fig:training_grid}(a). 
By monitoring computed validation errors, we adaptively and manually added training data to areas with large validation errors.
In this paper, we mainly showcase the results for the training data with total $N_{\mathrm{train}} = 275$ [see Fig.~\ref{fig:training_grid}(b)].
In Appendix~\ref{sec:initial_grid}, we analyze the effect of the adaptive refinement of the training grid in more detail.

\subsection{Training conditions}\label{sec:training_condtions}
The loss function $L$ is defined as
\begin{align}
    L = \frac{1}{N_{\mathrm{train}}N_{\ibarw}}\sum_{t=1}^{N_{\mathrm{train}}}\sum_{k=1}^{N_{\ibarw}} \frac{(|\Sigma_{\mathrm{pred}}^{(t)}(\ibarw_k) - \Sigma_{\mathrm{train}}^{(t)}(\ibarw_k)|)^2}{(\mathrm{max_{\ibarw}}|\Sigma_{\mathrm{train}}^{(t)}(\ibarw)|)^2},
    \label{eq:loss}
\end{align}
where $N_{\mathrm{train}}$ denotes the number of training data points.
Moreover, $\Sigma_{\mathrm{train}}^{(t)}(\ibarw_k)$ and $\Sigma_{\mathrm{pred}}^{(t)}(\ibarw_k)$ represent the $t$th training data and prediction on the $k$th sampling Matsubara-frequency grid points, respectively.
We minimize Eq.~\eqref{eq:loss} using the Adam~\cite{kingma2014adam} optimizer with a starting learning rate of $10^{-5}$.
The learning rate is switched to $10^{-6}$ at half of the total epochs.
The training takes over 8 hours for 125,000 epochs and 275 training data points with a single NVIDIA A100 80 GB PCIe.

\section{Results}\label{sec:results}
In this section, we show the main results. 
In Sec.~\ref{subsec:main_result}, we show the results of the PINN model. 
In Sec.~\ref{sec:non-PINN}, we discuss the effect of the non-negative projection and Lehmann projection.

\subsection{Results of the PINN model}\label{subsec:main_result}

\begin{figure}
    \centering
    \includegraphics[width=1\linewidth]{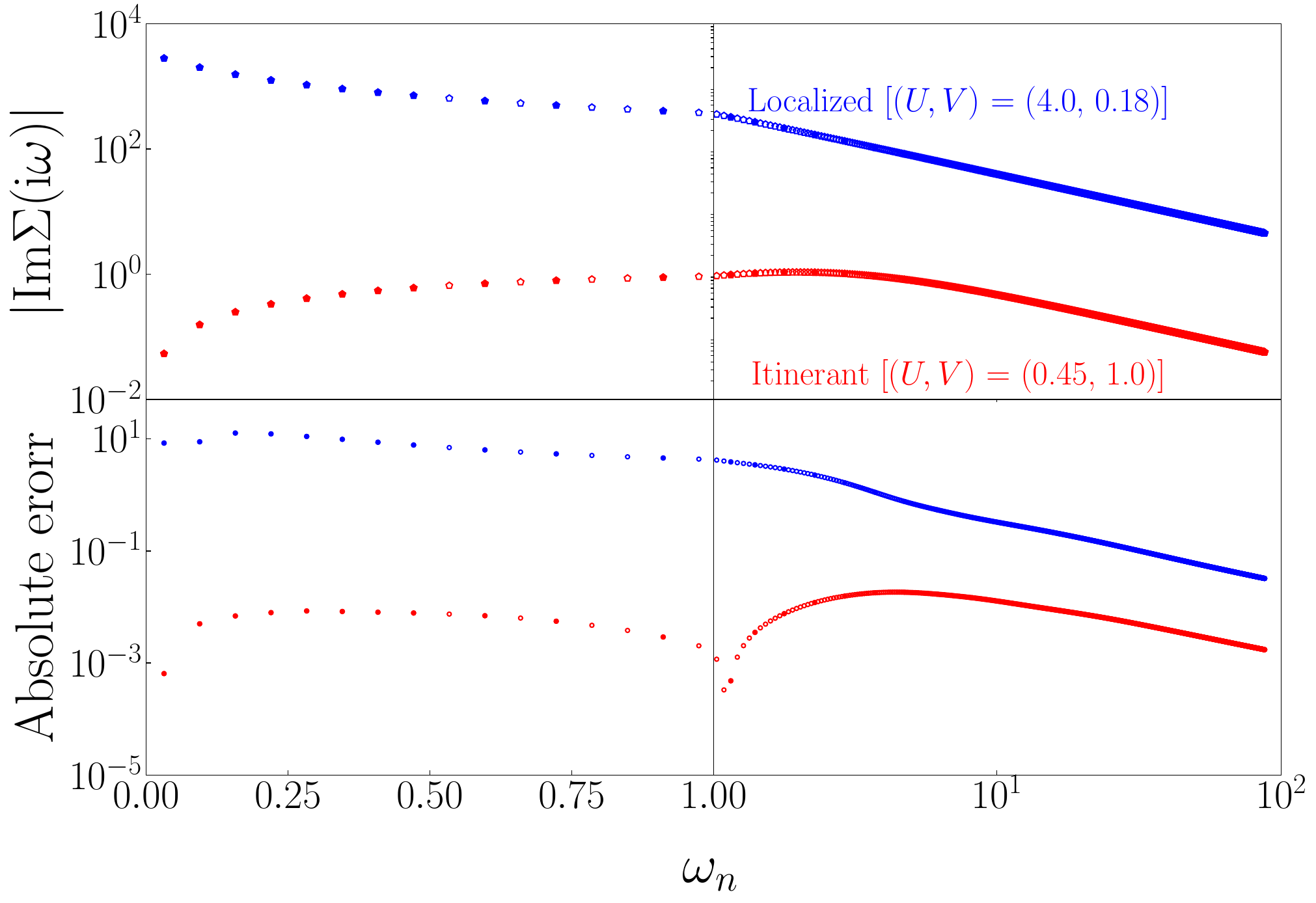}
    \caption{The self-energy as a function of Matsubara frequency $\w_n$ for typical itinerant and localized parameters.
    The upper panel shows the predicted values. 
    The lower panel displays the absolute errors between the test data and predicted values.
    The filled markers denote values at sparse Matsubara-frequency grid points, and unfilled markers indicate values at Matsubara-frequency grid points.
    }
    \label{fig:omegadep_sigmaiw}
\end{figure}

We show the main results of the present study in Fig.~\ref{fig:omegadep_sigmaiw}. 
Figure~\ref{fig:omegadep_sigmaiw} shows the predictions of the self-energy and absolute errors between predictions and test data for typical parameters in the itinerant regime $[(U, V)=(0.31, 1.0)]$ and in the localized regime $[(U, V)=(4.0, 0.15)]$ for $\beta=100$. 
As shown in the upper panel of the figure, the PINN model accurately predicts the self-energy at high Matsubara frequencies.
We will conduct a more quantitative analysis on the high-frequency data in Sec.~\ref{sec:non-PINN}.

As mentioned in the introduction, it is more relevant to focus on the predictions at low frequencies rather than at high frequencies.
The predictions accurately reproduce the behavior at low Matsubara frequencies: in the itinerant regime, the values linearly approach zero, while in the localized regime, they grow.
Furthermore, the absolute error at $\iw_0$ in the itinerant parameters is as small as below 0.01.
How this error propagates into physical quantities such as $Z$ will be discussed later.

\begin{figure}
    \includegraphics[width=\linewidth]{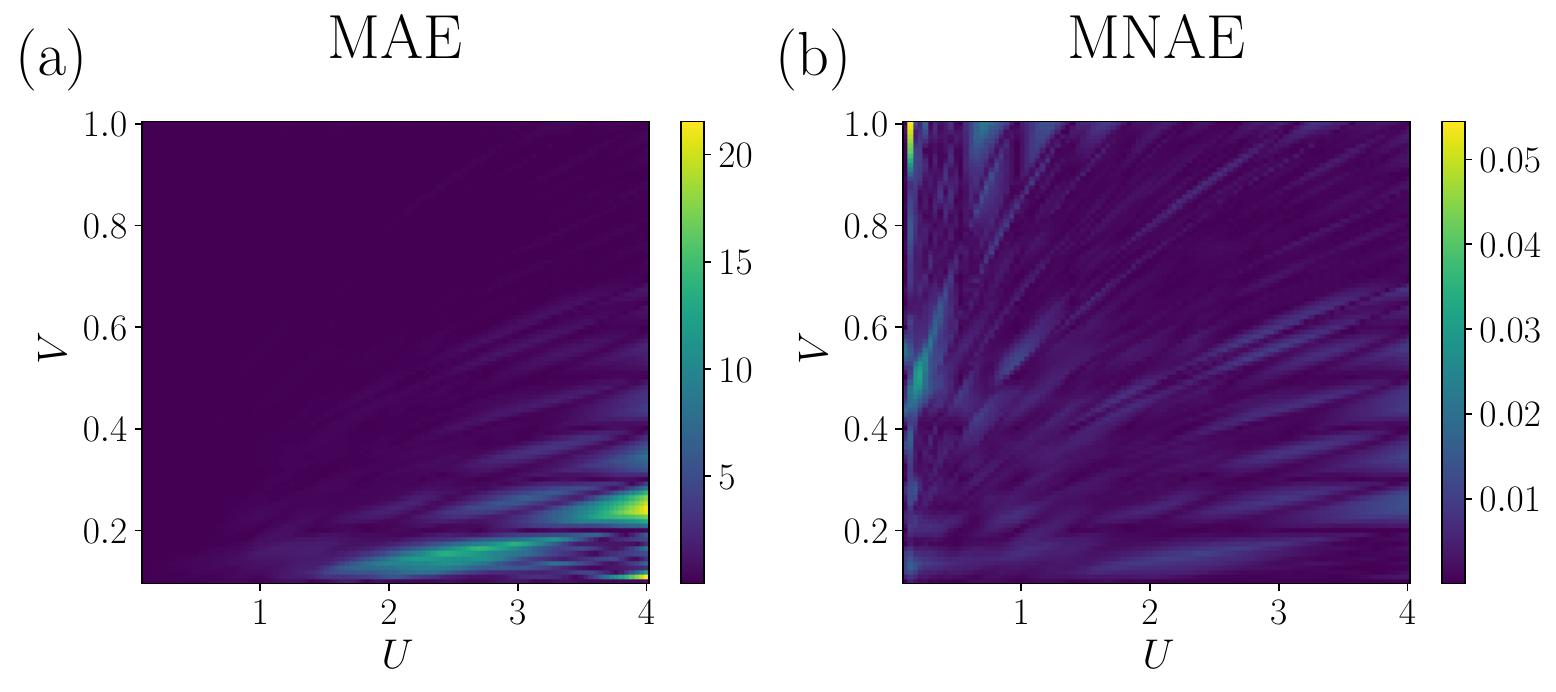}
    \caption{$U$-$V$ map for (a) MAE and (b) MNAE for the same PINN model as in Fig.~\ref{fig:omegadep_sigmaiw}.
    }
    \label{fig:sigmaiw_erroemap}
\end{figure}

Figure~\ref{fig:sigmaiw_erroemap} displays the mean absolute error (MAE) and mean normalized absolute error (MNAE) in the $U$-$V$ space.
On $(i, j)$ grid point $[(U, V) = (U_i, V_j)]$, the MAE $E_{\mathrm{MAE}}^{(i, j)}$ is defined by
\begin{align}
    E_{\mathrm{MAE}}^{(i, j)} = \frac{1}{N_{\ibarw}}\sum_{k=1}^{N_{\ibarw}}|\Sigmapred^{(i, j)}(\ibarw_k) - \Sigmatest^{(i, j)}(\ibarw_k)|, 
\end{align}
and the MNAE is defined by $E_{\mathrm{MNAE}}^{(i, j)}$
\begin{align}
    E_{\mathrm{MNAE}}^{(i, j)} = \frac{\sum_{k=1}^{N_{\ibarw}}|\Sigmapred^{(i, j)}(\ibarw_k) - \Sigmatest^{(i, j)}(\ibarw_k)|}{N_{\ibarw}\mathrm{max}_{\ibarw}|\Sigmatest^{(i,j)}(\ibarw)|}.
\end{align}
Here, $\Sigmatest^{(i, j)}(\ibarw)$ and $\Sigmapred^{(i, j)}(\ibarw)$ represent the output of the test data and prediction from the input of the test data corresponding to $(i, j)$ grid point in $U$-$V$ space.
Figures~\ref{fig:sigmaiw_erroemap}(a) and \ref{fig:sigmaiw_erroemap}(b) demonstrate that the predictions are highly accurate across a wide range of the $U$-$V$ plane. 
Now, we focus on MNAE, which holds large values in the itinerant parameter regime. 
In Fig.~\ref{fig:sigmaiw_erroemap}(b), it is seen that large values are found within the range of around $U=0.1$ and $V\approx0.9$ to $1.0$, and the range of $U=0.1$ to $0.4$ and $V=0.4$ to $0.6$.
These regimes match the regimes where adaptively added grid points are less or absent, as illustrated in Fig.~\ref{fig:training_grid}(b).
This indicates that adding more grid points to this regime could further reduce MNAE.
In Appendix~\ref{sec:initial_grid}, we compare the learning results using the training data from the initial grid and the adaptive grid and demonstrate that the regimes with large errors in the physical quantity $Z$, which is similar to the MNAE, are reduced by the training data.

\begin{figure*}
    \centering
    \includegraphics[width=0.8\linewidth]{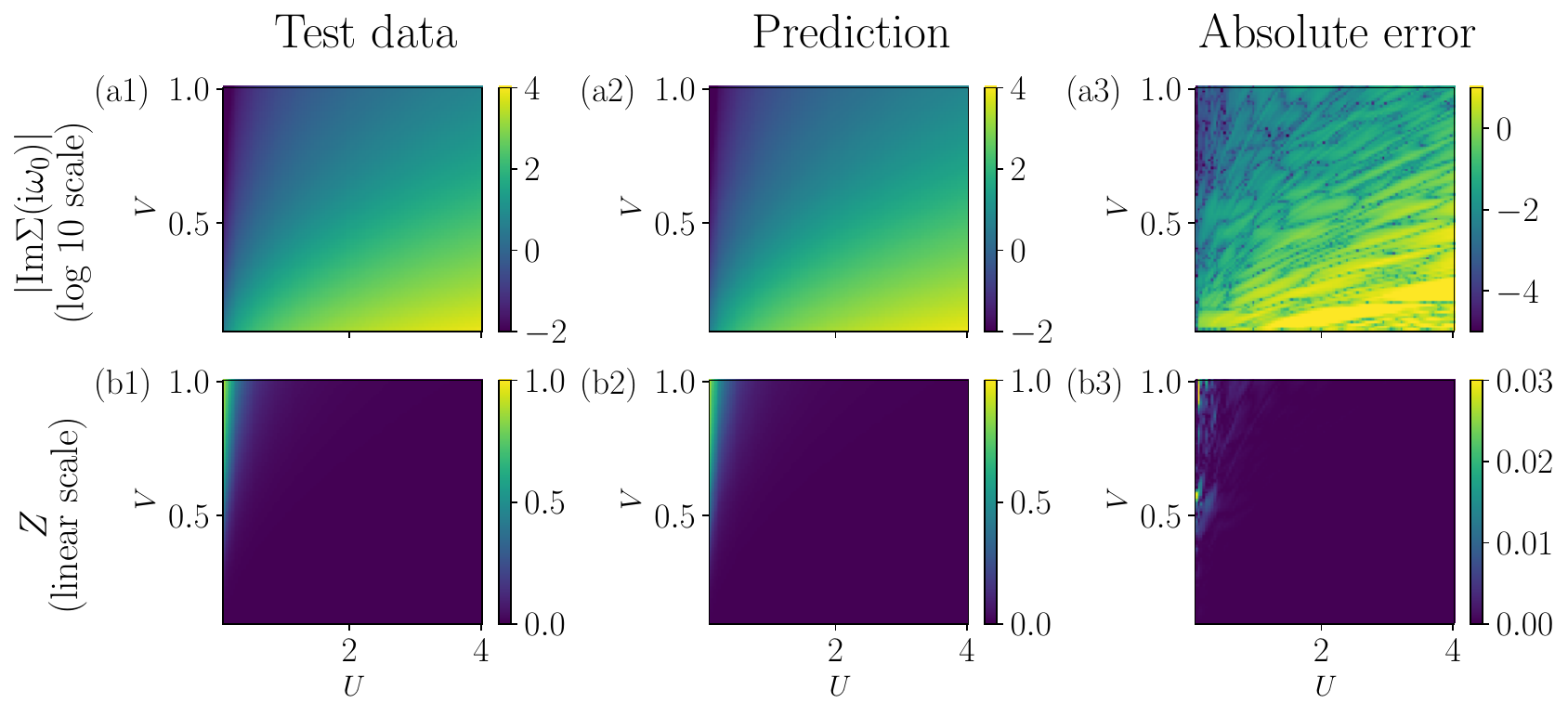}
    \caption{$U$-$V$ map of $\abs{\Im\Sigma(\iw_0)}$ on a logarithmic scale with base 10 [(a1)--(a3)] and $Z$ on a linear scale [(b1)--(b3)].
    Left panels: the test data. 
    Middle panels: prediction of the trained model. 
    Right panels: the absolute error.  
    }
    \label{fig:sigmaiw0_Z_adaptivegrid}
\end{figure*}

To test the PINN model in more depth, we evaluate $\abs{\Im\Sigma(\iw_0)}$ and the renormalization factor $Z$ defined by
\begin{align}
    Z =  \frac{1}{1 - \eval{\pdv{\Re{\Sigma(\omega)}}{\omega}}_{\omega=0}},
    \label{eq:Z}
\end{align}
on every $U$-$V$ grid point in the test data set.
In the DMFT context, the renormalization factor $Z$ is proportional to the ratio $m^*/m$ in lattice systems and characterizes an important low-frequency property of the system, where $m^*$ represents an electron effective mass and $m$ is an electron mass.
For example, $Z$ vanishes in a Mott localized regime for a half-filled single-orbital Hubbard model, while in the itinerant regime, $Z$ becomes nonzero and increases as $U$ decreases.
In the present study, we compute $Z$ as
\begin{align}
   Z \simeq \frac{1}{1 - \frac{2\Im \Sigma(\iw_0)}{\omega_0}}.
   \label{eq:Z_approx}
\end{align}

Figure~\ref{fig:sigmaiw0_Z_adaptivegrid} shows the computed results of both quantities.
As mentioned, the values of $\abs{\Im\Sigma(\iw_0)}$ are large in the localized parameter regime.
Even in the $U$-$V$ regime, where the values of the exact data are as large as $10^4$, the absolute error is no more than about 10.

The lower panels of Fig.~\ref{fig:sigmaiw0_Z_adaptivegrid} show the results of $Z$. 
As shown in Fig.~\ref{fig:sigmaiw0_Z_adaptivegrid}(b2), $Z$ is large in the itinerant parameter regime and increases as $V$ increases.
The errors are basically small but slightly large in the regime where $U=0.1$ to $0.4$ and $V=0.5$ to $0.6$. 
As previously mentioned, the regime with smaller errors contains many training data points, while the area with more significant errors has comparatively less.
These findings suggest that we could improve the accuracy of the PINN model by adaptively adding more test data points.

\begin{figure}
    \centering
    \includegraphics[width=\linewidth]{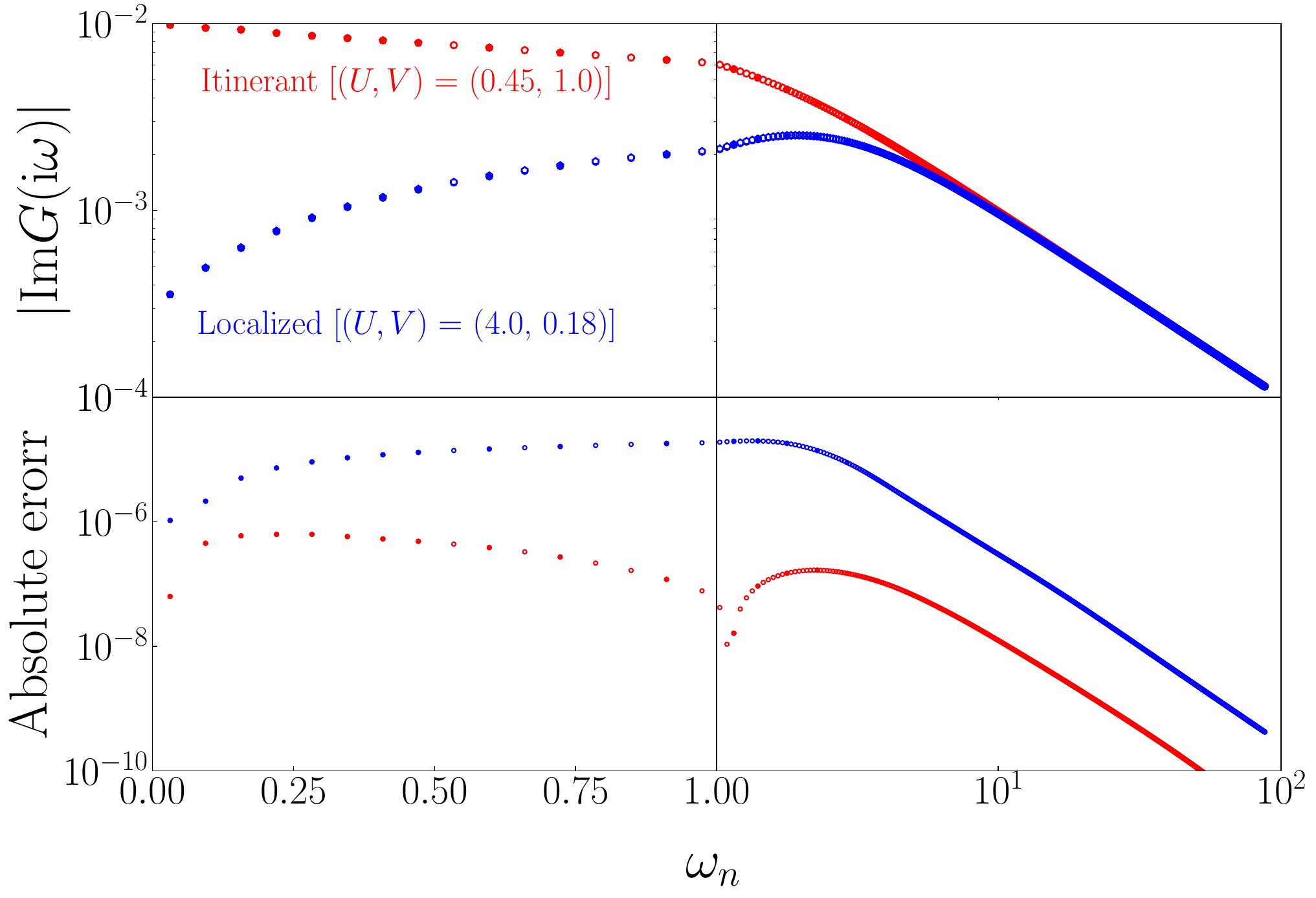}
    \caption{Green's function as a function of Matsubara frequency $\w_n$ for typical itinerant and localized parameters.
    The upper panel shows the predicted values. 
    The lower panel displays the absolute test errors.
    The filled markers denote values at sparse Matsubara-frequency grid points, and unfilled markers indicate values at Matsubara-frequency grid points.
             }
    \label{fig:omegadep_giw}
\end{figure}

We now evaluate the local Green's function by solving the local Dyson equation~\eqref{eq:dyson_eq} from the predicted self-energy.
Figure~\ref{fig:omegadep_giw} shows the results of the evaluated Green's function for the same $(U, V)$ pairs as in Fig.~\ref{fig:omegadep_sigmaiw}.
As shown in the figure, we can precisely predict Green's function in both regimes.

\subsection{Effectiveness of non-negative constraint and Lehmann projection} \label{sec:non-PINN}
\begin{figure*}
    \centering
    \includegraphics[width=0.8\linewidth]{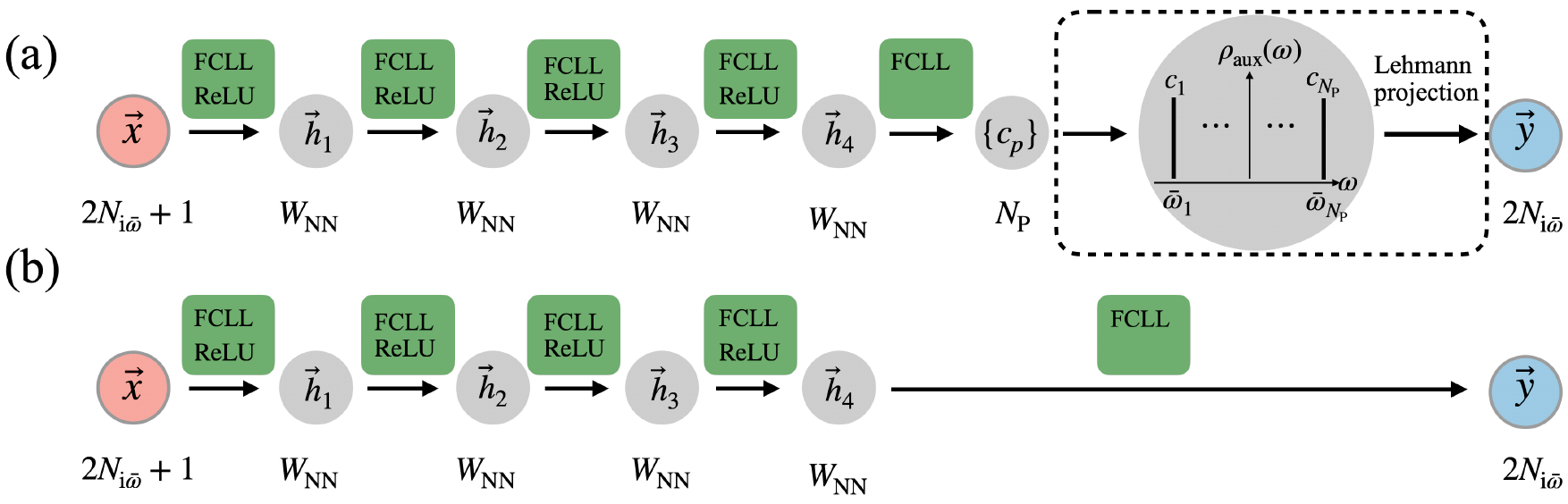}
    \caption{Schematic of non-PINN models.
             (a) NN model without the non-negative projection.
             (b) NN model without Lehmann projection.
    }
    \label{fig:NNmodel_variant}
\end{figure*}

Next, we explore the effects of utilizing physical constraints in our PINN model.
To this end, we use two variants of our model with reduced physics information, non-PINN models, which are illustrated in Figs.~\ref{fig:NNmodel_variant}(a) and \ref{fig:NNmodel_variant}(b), respectively.
The first one omits the non-negative projection, thereby lifting the non-negative constraint on the auxiliary spectral function of the self-energy $\rho_{\mathrm{aux}}(\w)$.
For the second one, we replace the Lehmann layer, including the ReLU layer, with the non-negative projection by the FCLL layer.

We trained these models using the same adaptive training grid with a sufficiently large number of epochs: 400,000 epochs for the NN model without the non-negative projection and 1,000,000 epochs for the one without the Lehmann projection.

To evaluate the accuracy of the NN models at high and low frequencies, we discuss two typical corresponding quantities. 
The first is an electron filling $\expval{n_{\sigma}}$, used to assess accuracy at high frequencies, calculated as 
\begin{align}
   \expval{n_{\sigma}}  = \sqrt{-\frac{\Sigma(\tau=0^+) + \Sigma(\tau=\beta^-)}{U^2}}.
   \label{eq:electren_filling}
\end{align}
The second quantity is the renormalization factor $Z$, used to assess accuracy at low frequencies, defined by Eq.~\eqref{eq:Z_approx}.

\begin{figure*}
    \centering
    \includegraphics[width=0.8\linewidth]{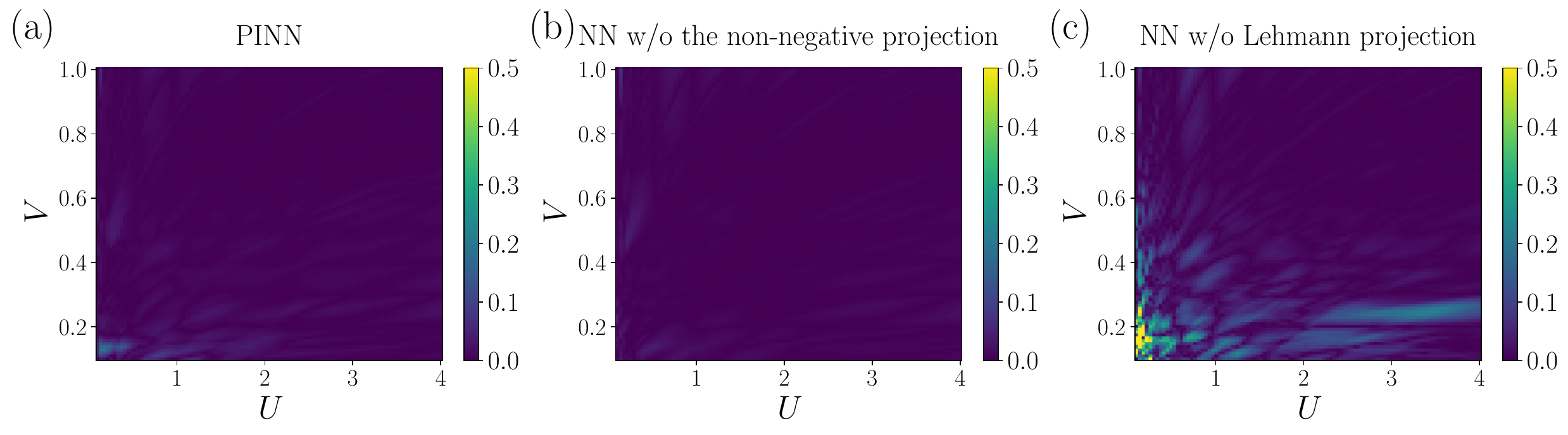}
    \caption{Effectiveness of the non-negative projection and Lehmann projection on the estimation of electron filling $\expval{n_{\sigma}}$. $U$-$V$ maps showing the absolute test error in $\expval{n_{\sigma}}$ for the following models:  
    (a) PINN model,
    (b) NN model without the non-negative projection,
    (c) NN model without Lehmann projection.
    All NN models are trained on the same $U$-$V$ grid points, as depicted in Fig.~\ref{fig:training_grid}(b).
    }
    \label{fig:cmap_filling_abserror_variantNN}
\end{figure*}

\begin{figure*}
    \centering
    \includegraphics[width=0.8\linewidth]{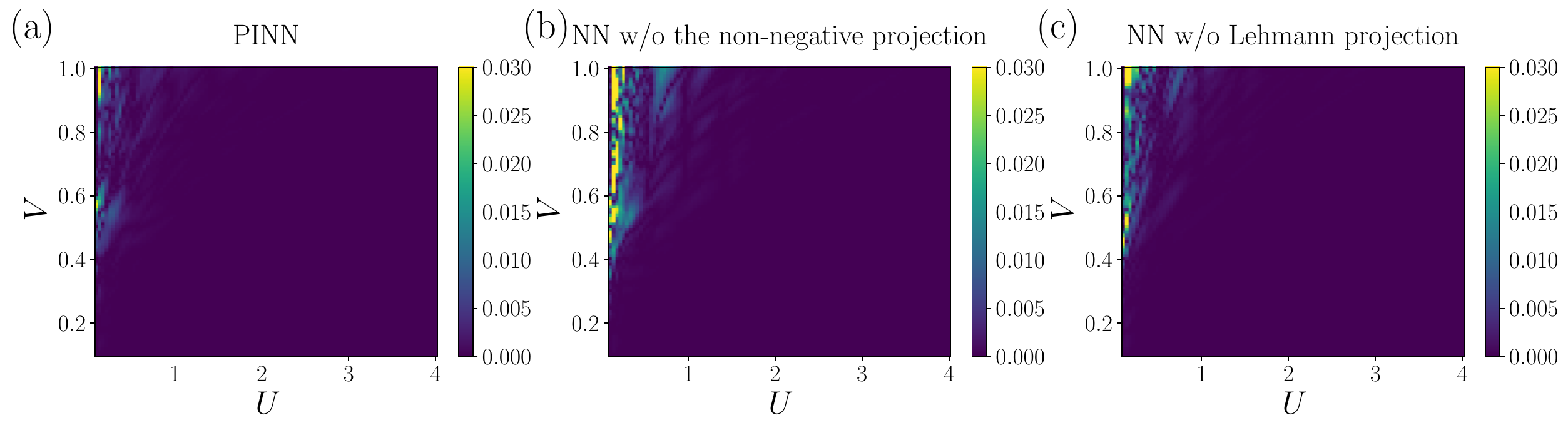}
    \caption{Effectiveness of the non-negative projection and Lehmann projection on the estimation of the renormalization factor $Z$. $U$-$V$ maps showing the absolute test error in $Z$ for the following models: 
    (a) PINN model,
    (b) NN model without the non-negative projection,
    (c) NN model without Lehmann projection.
    All NN models are trained on same $U$-$V$ grid points shown in Fig.~\ref{fig:training_grid}(b)
    }
    \label{fig:cmap_Z_abserror_variantNN}
\end{figure*}

Figures~\ref{fig:cmap_filling_abserror_variantNN} and \ref{fig:cmap_Z_abserror_variantNN} summarize the $U$-$V$ maps of the absolute test errors in $\expval{n_{\sigma}}$ and $Z$ predicted by the original PINN model, the NN model without the non-negative projection and the NN model without Lehmann projection.

First, we examine the effectiveness of the non-negative constraint on the auxiliary spectral function of the self-energy.
Figure~\ref{fig:cmap_filling_abserror_variantNN}(b) shows test errors in $\expval{n_{\sigma}}$ for the NN model without the non-negative projection.
The overall test errors are lower than those of the original PINN model, indicating that the NN model without the non-negative projection can predict the self-energy at high frequencies with slightly greater accuracy.
However, as shown in Fig.~\ref{fig:cmap_Z_abserror_variantNN}(b), the test errors in $Z$ remain large over a broader parameter range compared to that of the PINN model, specifically from $U\approx0.1$ to $U\approx1.0$ and $V\approx0.4$ to $V\approx0.5$.
These suggest that the non-negative constraint contributes to balanced generalization performance across both high and low frequencies.

We now discuss the effect of utilizing the Lehmann projection.
Figure~\ref{fig:cmap_filling_abserror_variantNN}(c) shows significant large test errors of $\expval{n_{\mathrm{\sigma}}}$ over the range of $U=0.1$ to $U\approx4.0$ and $V=0.1$ to $V\approx0.65$, covering a much broader error range compared to the other two NN models. 
The maximum test error is approximately 1.5, which is approximately eight times larger than that of the PINN model.
Next, we show the result for $Z$ in Fig.~\ref{fig:cmap_Z_abserror_variantNN}(c). 
The test errors are generally larger than those of the PINN model. including the range of $V\approx0.5$ to $V\approx0.6$ and $U\approx0.1$ to $U\approx0.3$.
These indicate that utilizing Lehmann representation improves the generalization performance at both high and low Matsubara frequencies remarkably.

\section{Summary}\label{sec:summary}
In this study, we proposed the PINN model based on the Lehmann representation and sparse modeling techniques, specifically designed to predict the self-energy of the SAIM.
We tested the PINN model using the SAIM for a noninteracting bath with semicircular DOS. 
We demonstrated that our PINN model accurately predicts the self-energy and other physical quantities calculated from it.

To get deeper insights, we explored the influence of the non-negative constraint on the auxiliary spectral function and Lehmann projection.
To this end, we developed two non-PINN models: one without the non-negative projection and another without the Lehmann projection. 
Our findings reveal that the non-negative constraint significantly enhances the balanced generalization performance of our PINN model at both high and low Matsubara frequencies.
Additionally, utilizing Lehmann representation improves the generalization performance at both high and low Matsubara frequencies remarkably.

Before concluding this paper, we discuss potential future directions for our research.
First, it is interesting to extend the PINN model from the particle-hole symmetric case to doped cases. 
A more important future direction is the application of the PINN model to self-consistent DMFT calculations.
Although the PINN model is applicable to general DOS scenarios, it is essential to extend the dataset to cover a higher-dimensional parameter space beyond the dataset used in this study.

\begin{acknowledgments}
F.K. and H.S. were supported by JSPS KAKENHI Grants No.18H01158, No.21H01041, No.21H01003, and No.23H03817, JST PRESTO Grant No. JPMJPR2012, Japan.
F.K. and H.S. thank Nobuyuki Yoshioka, Lei Wang, Xinyang Dong, Markus Wallerberger and Emanuel Gull for fruitful discussions.
We gratefully acknowledge the use of an exact diagonalization code developed by Markus Wallerberger, SparseIR.jl~\cite{WALLERBERGER2023101266} for sparse modeling techniques, and Flux.jl~\cite{Flux.jl-2018, innes:2018} for the development and training of NN models. 
The computation in this work has been done using the facilities of the Supercomputer Center, the Institute for Solid State Physics, the University of Tokyo.
\end{acknowledgments}

\appendix

\section{Convergence with the width of hidden layers}\label{sec:width_dep}
\begin{figure}
    \centering
    \includegraphics[width=1\linewidth]{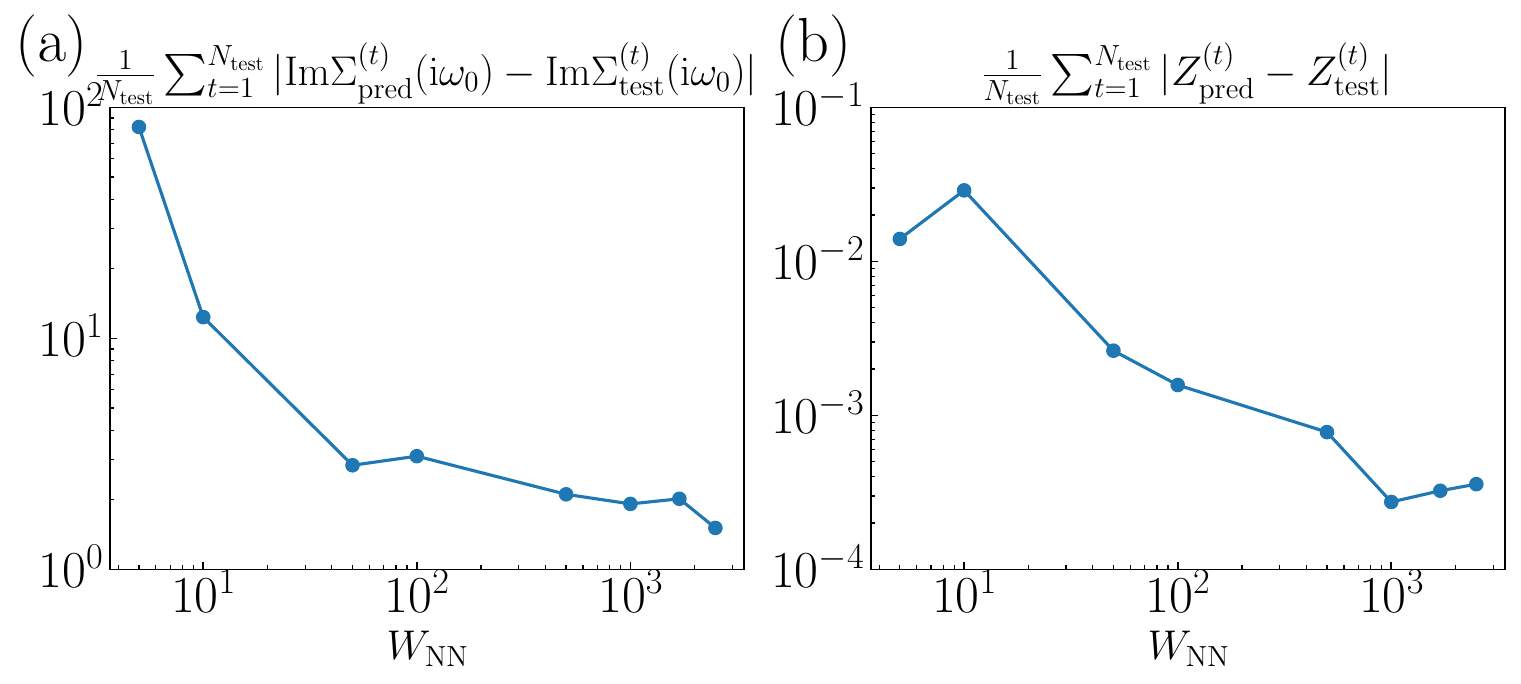}
    \caption{
    Effectiveness of increasing the width of each hidden layer $W_{\mathrm{NN}}$. 
    The averaged absolute error between predicted values and test data for (a) $\Sigma(\iw_0)$ and (b) $Z$ as a function of $W_{\mathrm{NN}}$.
    }
    \label{fig:width_dep}
\end{figure}

In this appendix, we examine the width dependence of the generalization performance of the PINN model trained on the adaptive grid shown in Fig.~\ref{fig:training_grid}(b).
Figures~\ref{fig:width_dep}(a) and \ref{fig:width_dep}(b) show the width dependence of the test errors for $\Sigma(\iw_0)$ and $Z$, respectively.
As shown in Fig.~\ref{fig:width_dep}(a), the absolute error in $\Sigma(\iw_0)$ decreases with increasing $W_{\mathrm{NN}}$ approximately following $1/W_{\mathrm{NN}}$ for $W_{\mathrm{NN}} \le 50$, and saturates for $W_{\mathrm{NN}} \ge 100$.
On the other hand, the absolute error in $Z$ decreases approximately following $1/\sqrt{W_{\mathrm{NN}}}$ for $W_{\mathrm{NN}} < 1000$ and saturates for $W_{\mathrm{NN}} \ge 1000$.
The saturation for $W_{\mathrm{NN}} \ge 1000$ indicates that increasing $W_{\mathrm{NN}}$ beyond this point does not improve the generalization performance of the PINN model.

\section{Effectiveness of the adaptive refinement of the training grid} \label{sec:initial_grid} 
\begin{figure}
    \centering
    \includegraphics[width=1\linewidth]{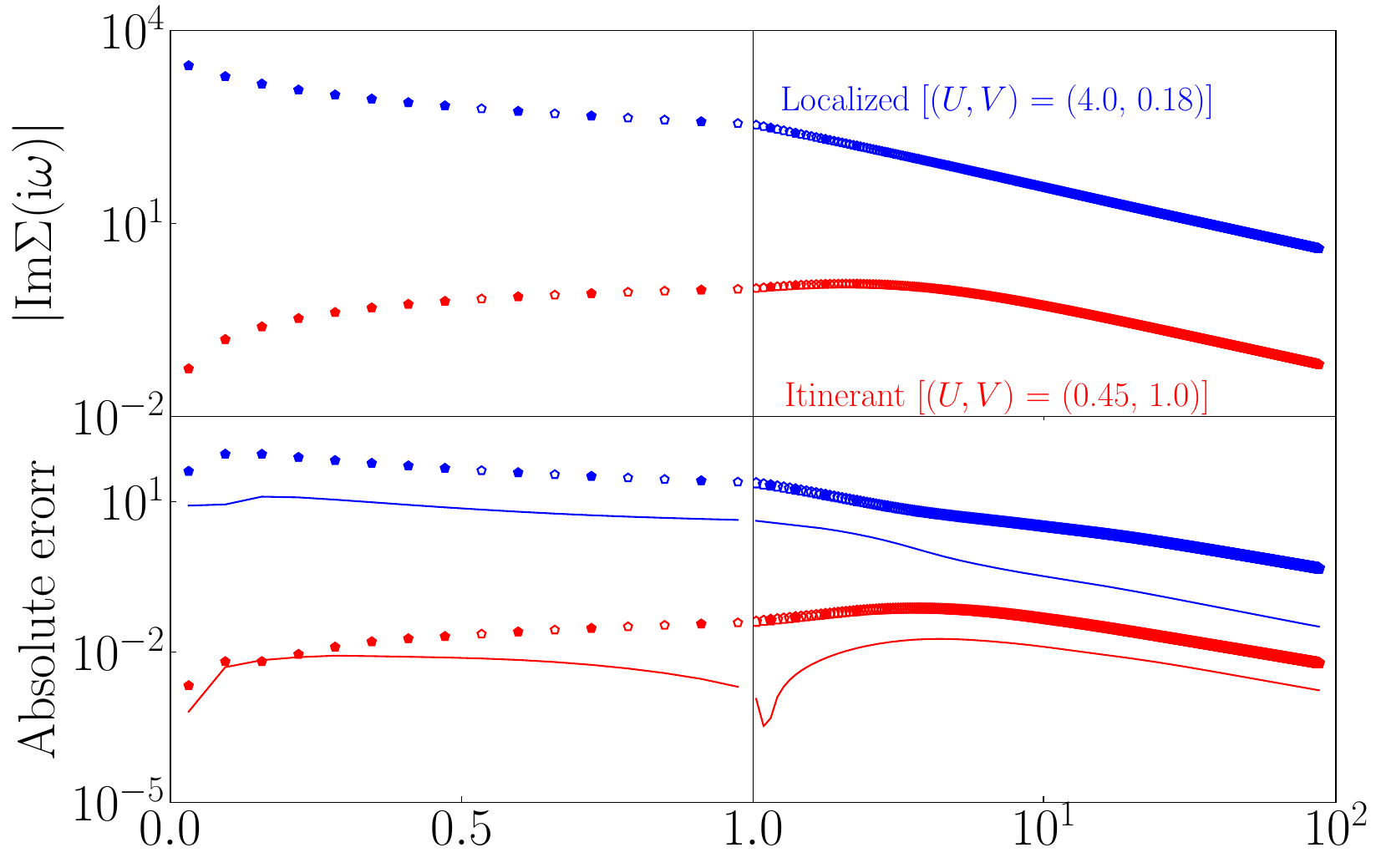}
    \caption{
    Effectiveness of the adaptive grid refinement.
    The self-energy as a function of Matsubara frequency $\w_n$ for typical itinerant and localized parameters.
    The upper panel shows the predicted values. 
    The lower panel displays the absolute test errors.
    Filled markers denote values at sparse Matsubara-frequency grid points, and unfilled markers indicate values at Matsubara-frequency grid points.
    In the lower panel, the absolute errors shown in Fig.~\ref{fig:omegadep_sigmaiw} are also plotted with lines for comparison.
    }
    \label{fig:sigmaiw_omegadep_initialgrid}
\end{figure}

\begin{figure*}
    \centering
    \includegraphics[width=0.8\linewidth]{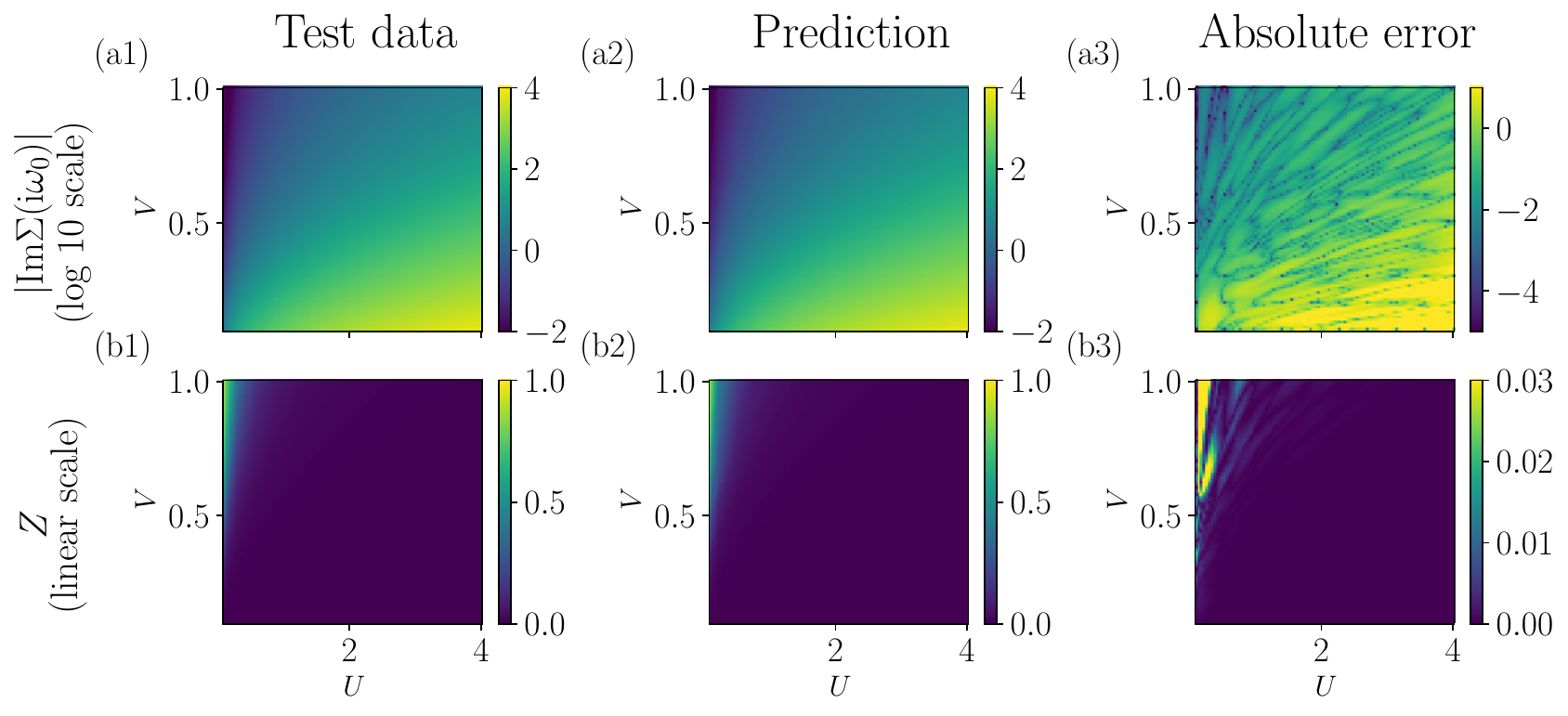}
    \caption{
    Effectiveness of the adaptive grid refinement. 
    $U$-$V$ map of $\abs{\Im\Sigma(\iw_0)}$ on a logarithmic scale with base 10 [(a1)--(a3)] and $Z$ in linear scale [(b1)--(b3)] using NN model trained on an initial $10\times 10$ $U$-$V$ grid.
    }
    \label{fig:cmap_sigmaiw0_Z_initialgrid}
\end{figure*}

In this appendix, we examine the influence of the adaptive grid refinement on the generalization performance.
Figure~\ref{fig:sigmaiw_omegadep_initialgrid} shows the $\w_n$-dependence of the self-energy estimated by the NN model trained on the initial grid for the same parameters for the data shown in Fig.~\ref{fig:omegadep_sigmaiw}.
The results reproduce the $\omega_n$-behavior qualitatively, while the absolute errors slightly worsen compared to those obtained by the NN model trained on the adaptive grid.
Figure~\ref{fig:cmap_sigmaiw0_Z_initialgrid} shows the $U$-$V$-dependence of $\abs{\Im\Sigma(\iw_0)}$, $Z$ and their absolute errors.
As shown in Fig.~\ref{fig:cmap_sigmaiw0_Z_initialgrid}(b3), the absolute error of $Z$ is larger than those obtained by the NN model trained on the adaptive grid across the regime where $U\approx0.1$ to $U\approx0.5$ and $V\approx0.5$ to $V=1.0$.
We conclude that the adaptive grid refinement improved the estimation accuracy of the PINN model in both itinerant and localized regimes.

\bibliography{main}

\end{document}